\documentclass {aa} 

\usepackage{times}
\usepackage{graphicx}

\begin{document}

\title{
Dust emission from 3C  radio galaxies and quasars:
New ISO\thanks{Based on observations with the Infrared Space Observatory
 ISO, an ESA project with instruments funded by ESA Member States
(especially France, Germany, the Netherlands and the United Kingdom)
and with the participation of ISAS and NASA.}
observations favour the unified scheme.
}
\author{
        K.\ Meisenheimer\inst{1}
   \and M.\ Haas\inst{1}
   \and S. A. H. \ M\"uller\inst{1,2}
   \and R.\ Chini\inst{2}
   \and U.\ Klaas\inst{1}
   \and D.\ Lemke\inst{1}
}

\offprints{Klaus Meisenheimer (meise@mpia-hd.mpg.de)}

\institute{
Max--Planck--Institut f\"ur Astronomie (MPIA), K\"onigstuhl \/17, 
\/D-69117 Heidelberg, Germany
\and
Astronomisches Institut, Ruhr-Universit\"at Bochum, \/D-44780 Bochum, Germany
}

\date{Received July 28, 2000 ; accepted February 13, 2001 }

\authorrunning{K. Meisenheimer {\it et al.}}
\titlerunning{3C radio galaxies and quasars}

\abstract{
In order to test the unified scheme for luminous radio galaxies and
quasars we observed 10 galaxy/quasar pairs from the 3CR catalogue with
ISOPHOT at infrared wavelengths between 5 and 180 $\mu$m.  Each pair
was selected such that both the 178 MHz luminosity and the redshift
match as close as possible between the radio galaxy and the quasar in
order to minimize effects of cosmic evolution. 13 of the 20 sources
were detected in at least one waveband. 
12 sources show clear evidence of a thermal bump at FIR
wavelength, while in the remaining 7 sources the upper limits
are still compatible with the presence of luminous dust emission.
In agreement with the predictions of the unified scheme, the quasars
and galaxies in our sample cannot be distinguished by their observed mid- and
far-infrared properties. \\
This is in contrast to the findings on the basis of the IRAS scans
which indicated that radio galaxies radiate significantly less mid- to
far-infrared emission than quasars.  However, the IRAS samples are
dominated by low-redshift sources ($z < 0.5$), while our sample
contains several of the most luminous radio galaxies at redshift $z
\simeq 1$. The latter have already been suspected to contain a hidden
quasar for other reasons, e.g. an extended emission line region
aligned with the radio axis. From the ratio between FIR luminosity
emitted by dust and the radio power at 178\,MHz, we conclude that the
radio galaxy/quasar unification might be perfectly valid for the most
luminous 3C sources at high redshift ($z \ga 0.8$). At lower redshifts
($z < 0.5$), however, some of the lobe-dominated FRII radio galaxies
contain active nuclei which emit less UV-optical continuum than the
quasars of similar radio power.  As this division is mainly a function
of redshift and less one of absolute radio power, we suggest that it
is caused by the evolution of the nuclear fueling rate with cosmic
epoch.  In order to quantify the deviation from the purely
aspect-dependent unified scheme at low redshifts a larger fraction of
3C radio galaxies has to be observed at mid- to far-infrared
wavelengths with sensitivities which suffice to yield secure
detections rather than upper limits.
}

\maketitle

\section{Introduction}

From an observational point of view, at least a dozen classes of
Active Galactic Nuclei (AGN) are discernible. Conceptionally,
however, there is widespread agreement that most, if not all, of
these different AGNs contain a massive black hole ($M_{bh} \ga
10^6$\,M$_{\sun}$ ) and that accretion onto this provides the main
source of energy. It seems natural, therefore, to explain their
differences by observational selection effects and thus restrict the
large variety of observed object classes. Such selection effects could
be differences in the evolutionary phase or the aspect angle which
might be important if AGNs are not spherically symmetric objects.

Indeed, already the earliest ``maps'' of Cygnus A (=\,3C\,405)
revealed the bi-polar morphology of the radio source (Jennison \& Das
Gupta, 1953). The detection of very bright spots of radio emission
(``hot spots'') in this double lobed radio source by Hargrave and Ryle
(1974), which can be connected by a straight line through the central
galaxy, led Scheuer (1974) and Blandford and Rees (1974) to the concept
of ``beams'' or ``jets'' which continuously feed the lobes with kinetic
energy. In the hot spots this energy is converted partly into
relativistic particles which then emit synchrotron radiation while
gyrating around magnetic field lines.  The jet model for the most
luminous radio galaxies (and quasars) is in itself highly
non-symmetric. But the full implications for the emitted radiation
pattern were only realized in 1977 when Scheuer and Readhead argued 
that the apparent super-luminal motions detected in the
milliarcsecond structure of some quasars (Whitney {\it et al.} 1971)
are caused by jet velocities close
to the speed of light. The Doppler boosting of the radiation from a
relativistic jet generates an extremely anisotropic radiation
pattern.  A few years later, Miller \& Antonucci (1983) detected a
broad emission line spectrum (i.e. a Seyfert I spectrum) in the
polarized (scattered) light from the Seyfert II galaxy NGC 1068,
pointing to the possibility that the appearance of thermal
radiation from AGNs also might be aspect dependent.

Many proposals have been put forward to unify observationally
distinct classes on the basis of different aspect angles ({\it i.e.}
the angle between the source axis and the line-of-sight).  Most
notable of these ``unified schemes'' are those for flat and steep
spectrum quasars (Orr \& Browne 1982), for Seyfert I and Seyfert II
galaxies (Antonucci \& Miller 1985), and for BL Lacertae Objects and
FRI radio galaxies (Urry \& Padovani 1995). Excellent reviews about
unified schemes and their limitations have been given by Lawrence
(1991) and Antonucci (1993).

The most radical proposal for the unification of radio-loud objects
was presented by Barthel (1989) who argued on the basis of the apparent
size distribution that the most luminous radio sources (that is the FR
II sources in the 3C catalogue) may represent an intrinsically
identical class of objects, only differing by the angle between the
radio axis and the line-of-sight. At its time, this was rather 
revolutionary in the
sense that two classes with vastly different optical appearance (some of the
brightest optical quasars and some of the faintest galaxies known
at that time) were postulated to be intrinsically the same.

A necessary ingredient in this radio galaxy/quasar unification is obviously not
only Doppler boosting of the non-thermal radiation from the jet but
also a non-spherical obscuration (analogous to that inferred for the
Seyfert II galaxies). If the jet axis and the plane of an obscuring
disk or ``torus'' ({\it i.e.} a disk with height similar to radius and
a central hole) are perpendicular to each other, the emergence of
(intrinsically) isotropic thermal radiation also can depend on the aspect
angle.  Support for this view came from the observations of the
so-called ``alignment effect'' in powerful, high-redshift radio
galaxies (Chambers {\it et al.} 1987, McCarthy {\it et al.} 1987), the emission
line regions of which are often very extended and closely aligned with
the radio axis. This could naturally be understood if a quasar hidden
by the torus obscuration from our line-of-sight provides the immense UV
luminosity which is required to excite the emission lines.

Conservation of energy implies that the ``hidden quasar light'' does
not just ``disappear'' in radio galaxies, but is converted into
another form of energy. In the most popular model, in which the quasar
is hidden by the dusty torus (on the line-of-sight $A_{\rm
V} > 50$\,mag is needed) the quasar light is absorbed by this torus and the
UV-optical radiation is heating the dust. The signature
of this warm dust is indeed well detected in the mid- and far-infrared
spectra of the brightest quasars (see e.g. Haas {\it et al.} 2000). Since
the thermal dust radiation (at temperatures between 30 and a few
hundred Kelvin) is emitted isotropically, and even the large optical
depths required for obscuring the optical light completely could
hardly suppress the re-radiation at $ \lambda > 20\,\mu$m, one expects
that the mid- and far-infrared thermal bump should provide an almost
orientation-independent measure of the core UV-optical luminosity (see Fig.\,1)

\begin{figure}
\begin{center}
\resizebox{8.8cm}{!}{\includegraphics {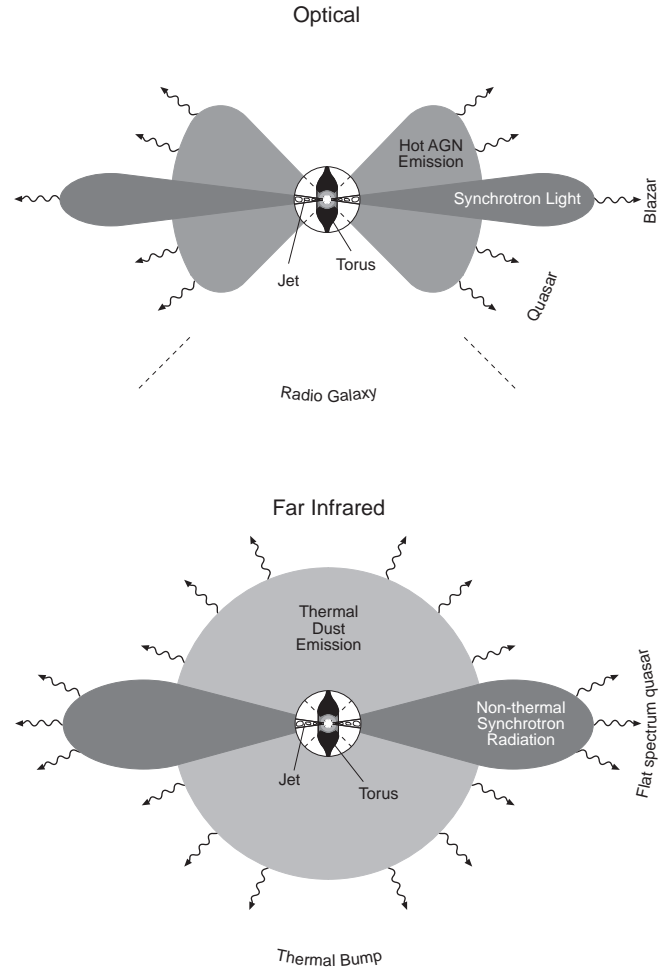} }
\end{center}
\caption[]{Comparison of the optical and FIR appearance of radio
galaxies and quasars at optical wavelengths ({\it top}) and in the
far-infrared ({\it bottom}). The inner circle contains the source
geometry with jets and a dusty torus. The radiation pattern
is sketched outwards. Note that the dusty torus surrounding the central engine
limits the angle under which the emission from the hot accretion disk
and the broad line region can be seen in the optical. Accordingly, an object will be
classified either as a radio galaxy or quasar. Objects seen pole-on show
in the optical the highly polarized and variable continuum
characterizing Blazars.}
\end{figure}

Thus, a direct test of the radio galaxies/quasars unification should
be possible when selecting a sample of radio sources on the basis of
an isotropically emitted radiation (e.g. the low frequency radiation
from the extended lobes, which gives a time average of the kinetic jet power
ejected from the core) and measuring their thermal core power by observing 
the dust emission between 10 and 200\,$\mu$m. If the unified scheme is
correct, then the objects classified as a radio ``galaxy'' must exhibit a
mid- and far-infrared bump comparable to those classified as quasars.
Note that the infrared radiation could be contaminated by the beamed 
synchrotron spectrum of flat-spectrum quasars which smoothly runs from the radio to the optical 
wavelength regimes. Thus the 
detection of a thermal bump above this synchrotron component is required.

The 3CR sample (Bennett 1962a,1962b) is selected at 178\,MHz, a
frequency at which most sources are dominated by the very extended
radio lobes. This sample, therefore, should suffer from the least
possible orientation bias. On the other hand it contains many rather
bright quasars which are within the reach of the sensitivity provided
by the ISOPHOT instrument on board of the Infrared Space
Observatory (ISO) and should therefore provide the best sample to test
the radio galaxy/quasar unification by its mid- to far-infrared properties.

First attempts to test the unified scheme along these lines have been
undertaken on the basis of the 12\,--\,100\,$\mu$m data from the IRAS
satellite.  Heckman {\it et al.} (1992) could detect only 6 out of 117 radio
galaxies and quasars in their sample of 3CR sources with $z > 0.3$
scanned by IRAS.  Since this number is too small to derive a
meaningful conclusion, they superimposed all IRAS maps of the radio
galaxies and quasars, respectively, thus constructing averaged
``super-maps'' of the IR emission from the two samples. On that basis
they conclude that 3C quasars on average are $4\times$ more luminous
at 60 to 100\,$\mu$m than a comparable sample of radio galaxies. This
is in pronounced contrast with the expectation from the unified
scheme. Later a similar conclusion was reached by Hes, Barthel \&
Hoekstra (1995) who found that only 6\% of the 3CR radio galaxies at
$0.3 < z < 0.8$ but 37\% of the quasars are detected on the finally
calibrated IRAS scans. Although this observational conflict with the
purely aspect-dependant unified scheme already has triggered several
theoretical models to produce a pronounced anisotropy of the mid- and
far-infrared emission ({\it e.g.} Pier \& Krolik, 1992, 1993), it
cannot be overlooked that the IRAS results are far from conclusive:
First, their wavelength coverage and sensitivity are not sufficient to
correct for the contribution by a (beamed) synchrotron component to
the FIR spectra of the quasars. 
Second, the ``super-maps'' could severely
be affected by fine-structure in the galactic cirrus such that one single
object located in a cirrus depression could alter the result of the whole sample. Therefore, it seems mandatory 
to reach sensitivities which suffice to detect individual sources
and the shape of their spectral energy distributions up to 200\,$\mu$m
before firm conclusions can be drawn.

Here we report a comparative study of 3CR radio galaxies and quasars
which was undertaken as part of the {\em ISO European Central Quasar
Programme} .  The pointed observations with ISOPHOT led to much deeper
detection limits than those reached by IRAS. New millimeter
observations with the IRAM 30\,m telescope help to determine the
synchrotron contribution at mid- to far-infrared wavelengths.  We
observed 10 radio galaxies and 10 quasars.  With 13 of these sources
detected we are for the first time in the position to compare radio
galaxies and quasars on the basis of individual detections, rather
than on the basis of sample averaged ``super-maps''. Some preliminary
results of our study have already been published (Haas {\it et al.} 1998).

\section{Sample selection and data base.}


For the {\em ISO European Central Quasar Programme} we have selected 
various samples of quasars.
In order to test the unified scheme for powerful radio galaxies and
quasars, we did not only select {\it quasars} from the 3CR
catalogue but included the same number of objects which 
have been classified as radio {\it galaxies} (optical identifications from 
Spinrad {\it et al.} 1985). The basic idea has been to observe
{\it pairs} of radio galaxies and quasars, the members of which are
closely matched in their extended lobe luminosity (as measured at
178\,MHz) and redshift in order to keep evolutionary effects on a
cosmic time scale in check. 

Our original sample contained 16 quasar/galaxy pairs in the redshift
range $0.15 < z < 2.1$.
When our observations with the chopped photometry mode of ISOPHOT
early in the mission led to
less satisfactory results, we decided that an optimum observation strategy
had first to be developed before the complete sample was observed. 
%
%
In the end only 17 sources of the original sample could be observed
with ISOPHOT. For 3C\,273, which was contained in our original sample,
we found an observation (by H. Smith) in the public domain of the ISO
Data Archive.  Two more radio galaxies (3C\,437, 3C\,405) were later
included in our list in order to replace other radio galaxies which
had dropped out of the visibility zone.  Thus our observed sample of
3CR sources contains 10 quasars and 10 radio galaxies, but only 3 of
the originally matched pairs. In order to match the remaining galaxies
and quasars as closely as possible along the original idea new pairs had
to be formed. Note that one original pair (3C368--3C287) has been
broken up in this process and we had to join the prominent sources
3C\,405 (Cygnus A) and 3C\,48 together although their redshift and
178\,MHz power are rather different (see Table 1).

\subsection{ISO Data} 

\begin{figure}
\vspace{0.5cm}
\begin{center}
\resizebox{7.5cm}{!}{\includegraphics {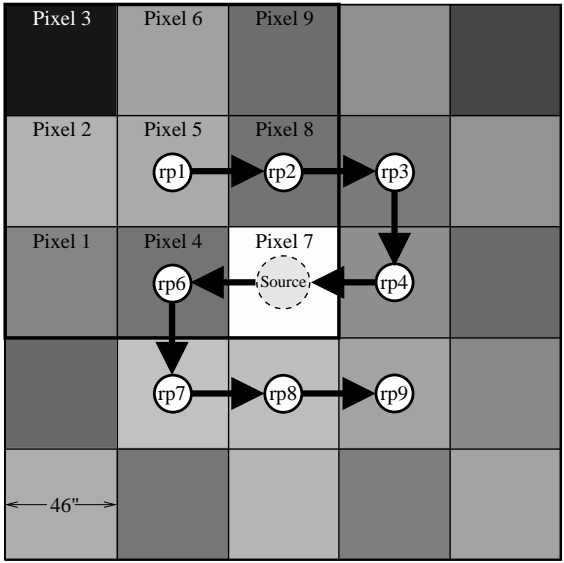} }
\end{center}
\caption[]{
Grey scale map of the faintest detection, 
3C380 at 60\,$\mu$m, using a small raster. 
The 3\,$\times$\,3 pixel detector array
(Pixel\,1, ..., Pixel\,9) 
is positioned at a 3\,$\times$\,3 raster point
grid (rp\,1, ..., rp\,9 denoting the central array positions)
with one pixel raster step size 
resulting in a 5\,$\times$\,5 pixel map. Thus each detector 
pixel hits the source exactly once. The nine-fold redundancy on-source 
as well as the low surrounding background structure provide the quality 
of the detection. (For this image no drift correction was applied, therefore 
the black pixel in the upper left corner with no redundancy has a lower flux 
level.)
}
\end{figure}
          
\begin{figure}
\resizebox{8.5cm}{!}{\includegraphics {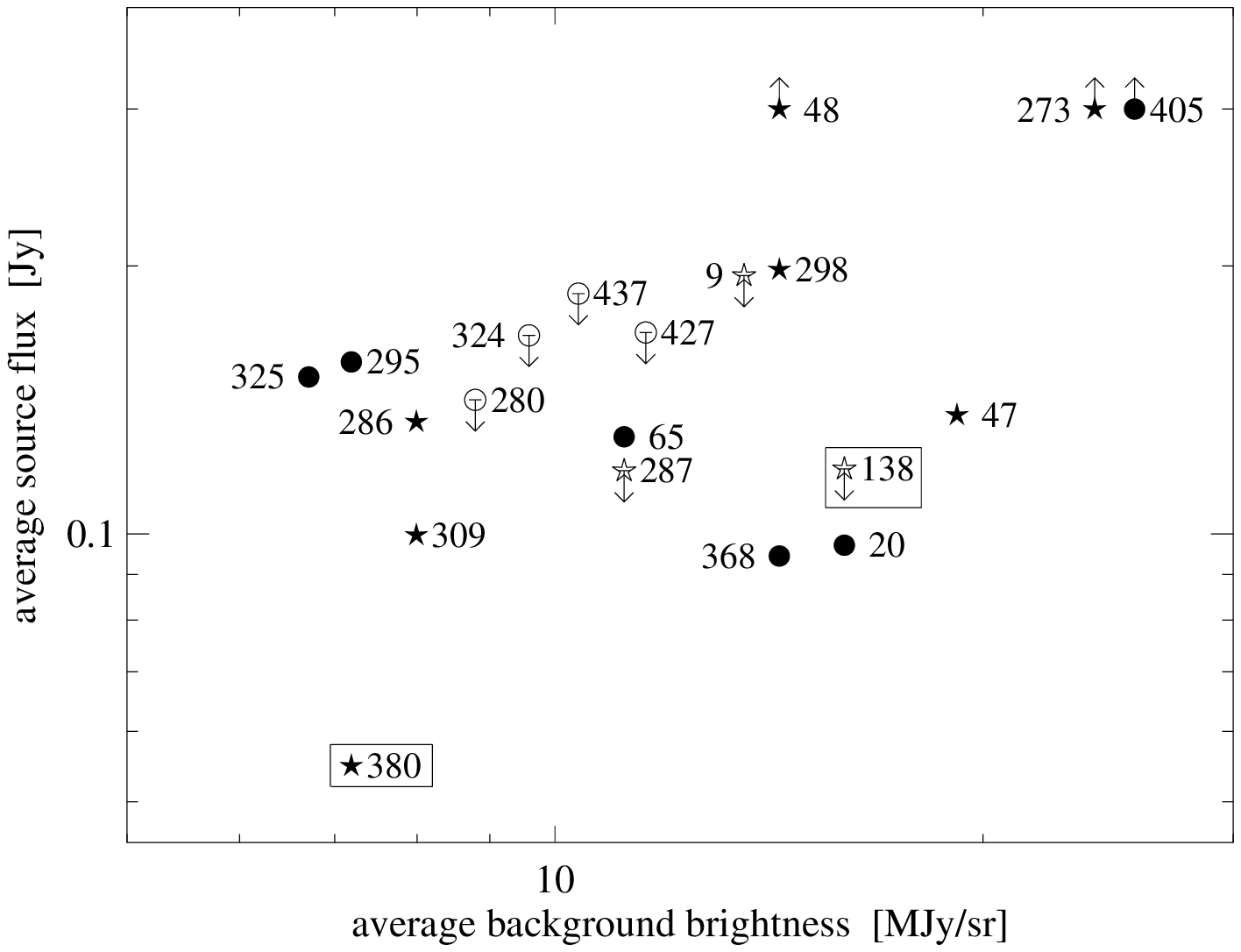} }
\resizebox{8.5cm}{!}{\includegraphics {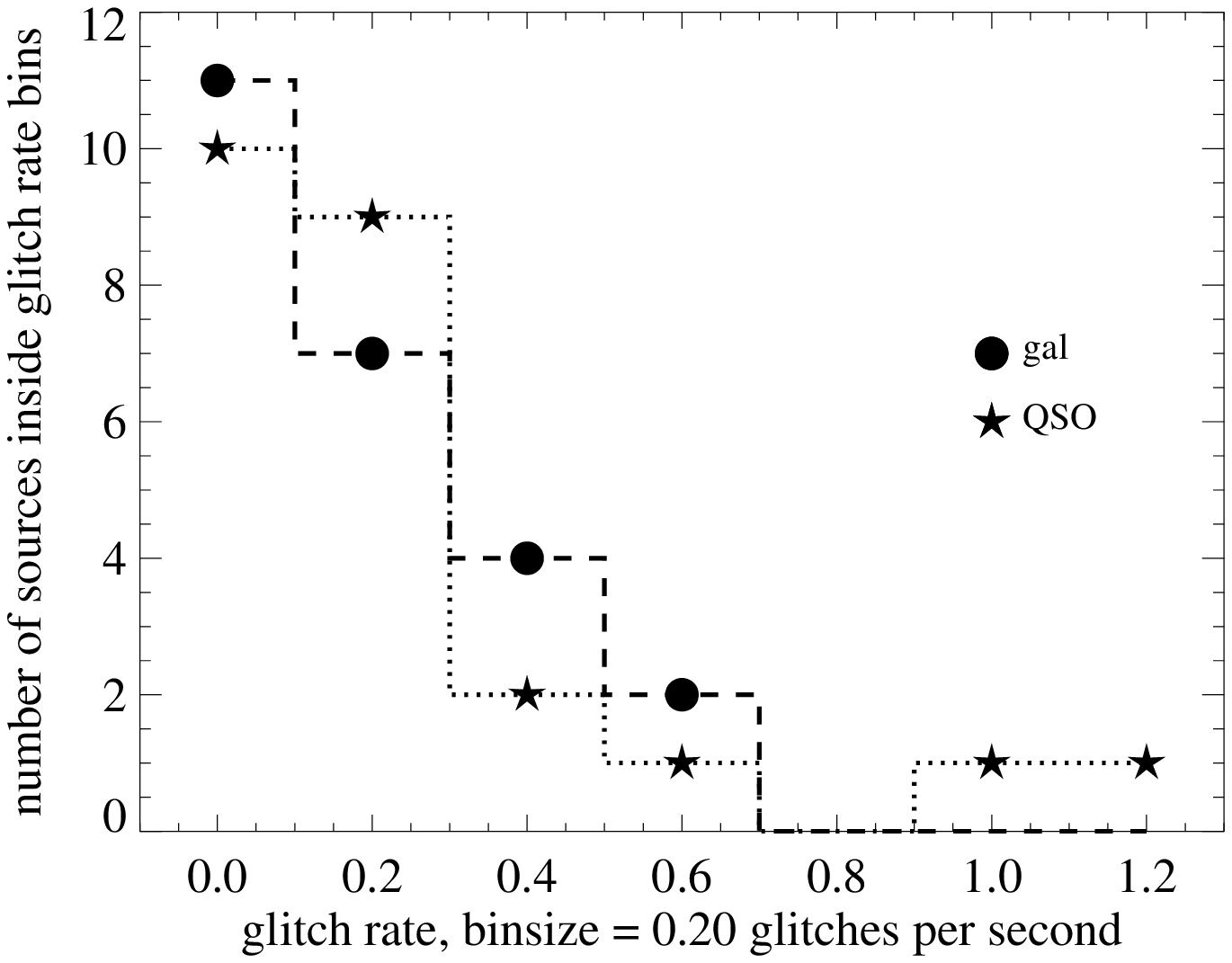}}
\caption[]{Illustration of the sample balance: The sensitivities of our
observations are not biased with respect to either quasars or radio
galaxies.\\ {\it Top:} Far-infrared source flux (averaged 60 and
100\,$\mu$m values) versus background brightness. The symbols are:
asterisks = quasars, circles = galaxies, filled = detections, open
symbols with arrow = 3--sigma upper limits. All data are observed in
chopped mode, except the two cases 3C380 and 3C138 marked with a
surrounding box for which more sensitive raster maps were
obtained. The three sources with fluxes well above 300 mJy are indicated
by upwardly pointing arrows.\\
{\it Bottom:} Histogram of the glitch rates among the 60 and 
100\,$\mu$m observations of quasars (asterisks) and radio galaxies (circles). 
Due to some repeated measurements the total number for quasars and 
for galaxies sums up to 24 instead of 20.
}
\end{figure}

The observations were performed with ISOPHOT (Lemke {\it et al.} 1996),  the
photometer on board ISO (Kessler {\it et al.} 1996). 
The observing  modes comprise chopped measurements and, adopted in 
the course of the ISO mission, small raster maps. 
The on-source integration time was chosen  between 16
and 360 s, depending on the source and cirrus brightness, and  the detector
and filter combination. The apertures used in flux derivation are 23$\arcsec$ 
circular  for
4.8--25$\mu$m (P1, P2 detectors), and 
46$\arcsec$ square for 60--100$\mu$m  (i.e. when chopping, then 
only the central pixel of the $3\,\times\,3$ pixels C100 array is used). For
120--180$\mu$m they are 92$\arcsec$ square in case of maps  (where the source
is centered on a pixel of the C200 array) and 184$\arcsec$ square in the case of chopped
observations  (where the source is centered in the mid of the 
$2\,\times\,2$ pixels C200 array).

The data were reduced using the Interactive 
Analysis tool (PIA V7.3 and V8.1) -- details and special refinements 
of the data analysis are described in Haas {\it et al.} (2000). 
The uncertainty of the absolute calibration 
is about 30\%.

\begin{figure}
\resizebox{8.5cm}{!}{\includegraphics {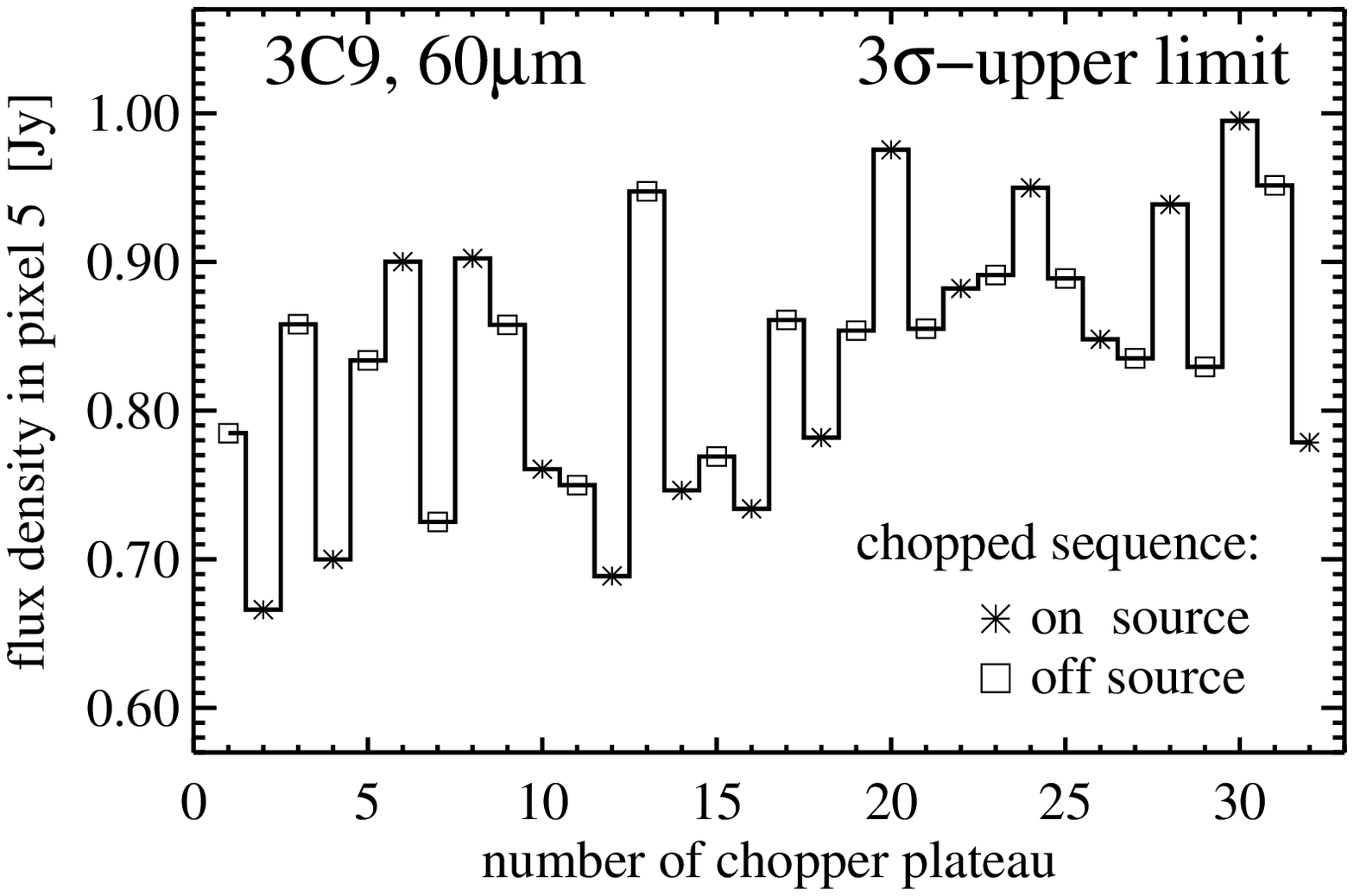} }
\resizebox{8.5cm}{!}{\includegraphics {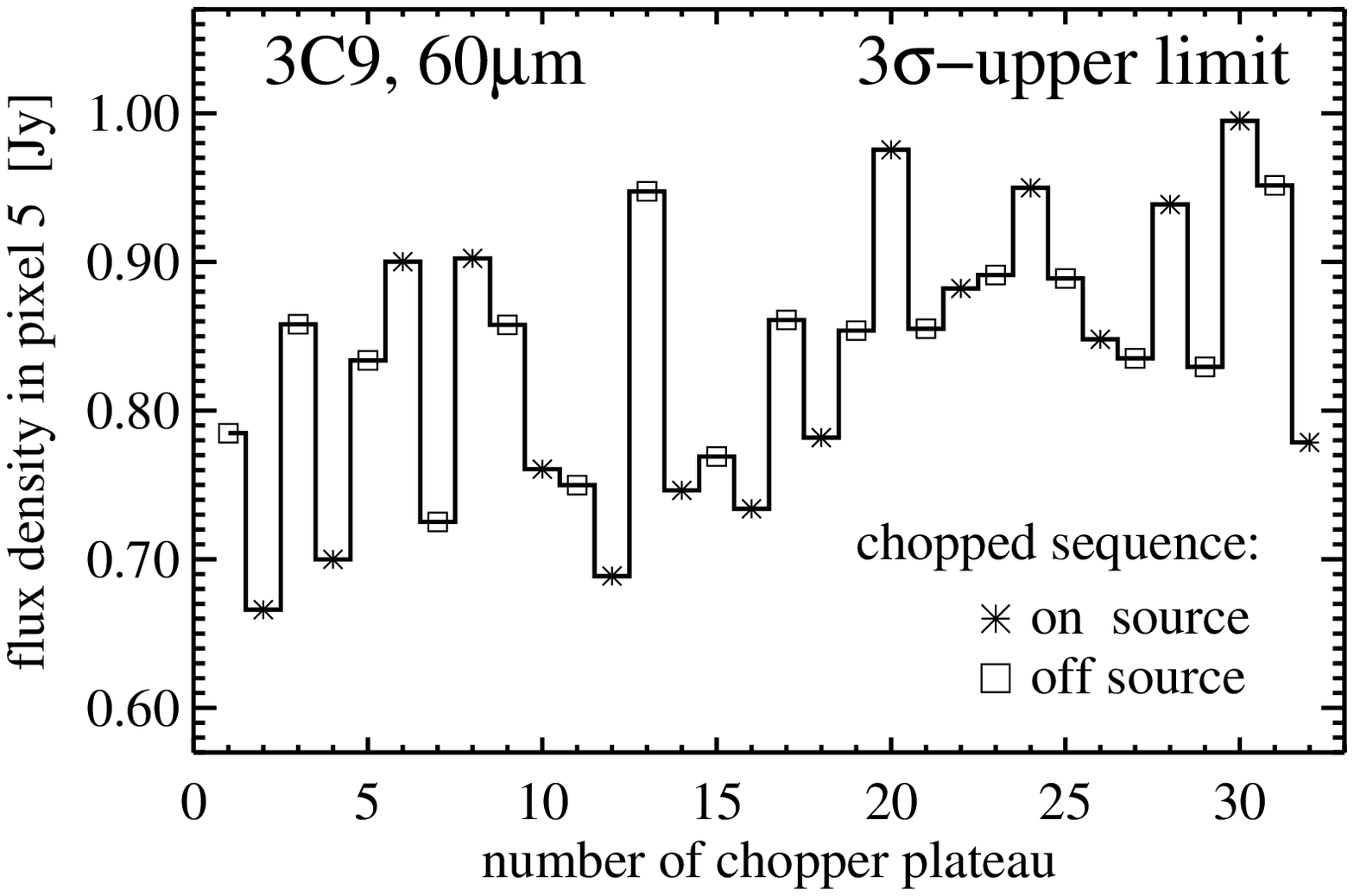} }
\caption[]{Influence of increased glitch rates on the source detection in 
chopped data sequences:\\
Chopped sequence of 3C9 and 3C298, both at similar average background 
levels, and of the same duration of 128 s, and with a long term detector 
drift. While the sequence of 3C298 clearly shows the on--off--source 
oscillation pattern typical for a clear detection with some remaining 
glitches at the chopper plateaux 5 -- 9, that of 3C9 is irregular and 
disturbed. Increased cosmic rates near the van Allen belt
have limited the detections at 60/100\,$\mu$m:
134/158 glitches for 3C9, but only 32/17 for 3C298.

}
\end{figure}
          
The observations were performed either in chopping mode or by
scanning mini-maps ({\it cf.} Table 1). For this comparative study it is
essential that the different observing modes do not bias the
detection statistics. 
Fortunately, we cannot find any obvious dependence between observing
mode and sensitivity:
For example, at 60 and 100\,$\mu$m 
the only two cases with maps are both on QSOs, but 
one of them (3C138) remains undetected, while one (3C380) which 
was not detected in the chopped mode, is now detected using the 
sensitive raster map (Fig.\,2). Although
this biases the statistics slightly in favour of quasars, 3C380 is 
by chance just the one case, where the FIR emission is clearly dominated by 
non-thermal synchrotron radiation (see below).  

Furthermore, detections may be hampered by bright cirrus 
as well as by cosmic particles hitting the detectors. 
Again, we cannot find any difference with respect to either galaxies or quasars. 
This is illustrated for the 60 and 100\,$\mu$m data: 

Fig.\,3 (top) shows source flux versus background brightness.
Galaxies and quasars are equally well distributed over the range 
of background brightness. 
At low background (less than 8.5 MJy/sr) all five sources are detected, 
two galaxies and three quasars (one being 3C380 with the lowest flux which is 
seen only on the raster map). 
At higher background non-detections are frequent. 
Remarkably, some chopped data (3C\,20, 3C\.47, 3C\,368) yielded rather
faint flux detections at a background brightness comparable to that
underneath 3C\,138, while for this source only
upper limits could be achieved on the raster maps.
Nevertheless, these detections seem reliable since their 60 to
100\,$\mu$m flux ratio implies a dust temperature well above typical
cirrus values.

The fact that a couple of upper limits are at quite high flux levels
could be caused by cosmic ray hits. Fig.\,4 shows the chopped sequence
of 3C\,9 and 3C\,298, both at the same background level. While the
sequence of 3C\,298 immediately shows the on--off oscillation
pattern typical for a clear source detection, that of 3C\,9 is irregular and
disturbed.  Obviously, increased cosmic rates have limited the
detections. 
%
We checked that the orbit
positions of our galaxy and quasar observations were randomly
distributed, and the
glitch rates 
show no significant bias with respect to either galaxies or quasars (Fig.\,3, bottom). 

\subsection{New millimeter data} 

In order to constrain the synchrotron contribution, 
additional 1.3\,mm continuum observations were obtained for 11
sources of our sample (see Table 1). They were carried out in several campaigns between 1996 
and 1999 at the IRAM 30\,m telescope on Pico Veleta, Spain, 
using the MPIfR 19 channel bolometer (Kreysa {\it et al.} 1998) 
in standard ON--OFF--OFF--ON beam switching mode.  
The atmospheric transmission was determined by measurement series at 
various zenith distances (sky dips); the calibration was
established by observations of Uranus. The absolute uncertainty of the 
1.3\,mm fluxes is estimated to be around 25\%.

\subsection{Supplementary data from the literature} 

Most of the literature flux values were found in the NED and SIMBAD data 
bases, in particular those for the IRAS detections. 
The optical fluxes are mainly compiled in Spinrad {\it et al.} (1985). 
The mm fluxes are measured by Steppe {\it et al.} (1988, 1992).
Further sub-mm data are found for 3C\,405 (Robson {\it et al.} 1998) and for 
3C\,324 (Best {\it et al.} 1998). 

Special monitoring including the years 1983 of the IRAS measurements and 
1996 of the ISO measurements was reported for 3C\,273 by  
Stevens {\it et al.} (1998), 
by T\"urler {\it et al.} (1999) and 
by Hans Ungerechts (IRAM, private communication).
Additionally, monitoring data at millimeter wavelengths are available
for 3C\,380 from Steppe {\it et al.} (1988, 1992).

\begin{table*}

\caption[]{ 
{\bf Measured flux density in mJy as a function of wavelength in $\mu$m.} 
   The sources are ordered along the pairs as shown in Fig.\,5.
   The ISOPHOT observing modes M (column 2) are: c = chopping with P1--C200,
   m = mapping with all detectors P1--C200, b = both: chopping
   with P1--C100 and mapping with C200 detetctors. 
   z is the redshift and  P$_{\rm 178}$ is the spectral radio power at
   178\,MHz in 10$^{\rm 26}$\,W\,/\,Hz  (for $H_0 = 75 {\rm
   kms^{-1}Mpc^{-1}}$ and $q_0 = 0$).
   Bold 3C numbers and flux values indicate ISO detections above
   the 3--sigma level. Non-detections are represented by thin numbers and
   are shown as $3\sigma$ upper limits in Fig.\,5.
   The last column gives the 60\,$\mu$m fluxes of the IRAS detections.
 }
\begin{tabular}{
        |l@{} r| c@{} r|
        r@{} c@{} r@{ }
        r@{} c@{} r@{ }
        r@{} c@{} r@{ }|
        r@{} c@{} r@{ }|
        r@{} c@{} r@{ }
        r@{} c@{} r@{ }|
        r@{} c@{} r@{ }
        r@{} c@{} r@{ }
        r@{} c@{} r@{ }|
        r@{} c@{} r@{ }|
        r@{} c@{} r@{}|
                         }
\hline
\multicolumn{2}{|l|}{\bf{Object}}       & 
\multicolumn{1}{|c}{\bf{z}}             & 
\multicolumn{1}{c|}{\bf P$_{\rm 178}$}  &  
\multicolumn{9}{|c|}{\bf P1-Detector}   & 
\multicolumn{3}{|c|}{\bf P2-Det.}       & 
\multicolumn{6}{|c|}{\bf C100-Detector} & 
\multicolumn{9}{|c|}{\bf C200-Detector} & 
\multicolumn{3}{|c|}{\bf IRAM} 		&
\multicolumn{3}{|c|}{\bf IRAS} 		\\
\multicolumn{1}{|l} {\bf 3C}            & 
\multicolumn{1}{r|} {\bf M}             & 
\multicolumn{1}{|c}{}                   & 
\multicolumn{1}{c|}{} 			&
\multicolumn{3}{|c} {\bf 4.8}           &
\multicolumn{3}{c}  {\bf 7.3}           &
\multicolumn{3}{c|} {\bf 12.8}          &
\multicolumn{3}{|c} {\bf 20}            & 
\multicolumn{3}{|c} {\bf 60}            & 
\multicolumn{3}{c|} {\bf 100}           &
\multicolumn{3}{|c} {\bf 120}           &
\multicolumn{3}{c}  {\bf 170}           & 
\multicolumn{3}{c|} {\bf 180}           & 
\multicolumn{3}{c|} {\bf 1300}          &
\multicolumn{3}{c|} {\bf 60}          	\\
\hline

{\bf 405}&b&0.056&552&1&$\pm$&4&{\bf 40}&$\pm$&{\bf 1}&
{\bf 250}&$\pm$&{\bf 36}&{\bf 590}&$\pm$&{\bf 59}
&{\bf 3034}&$\pm$&{\bf 56}&{\bf 2155}&$\pm$&{\bf 57}
&{\bf 994}&$\pm$&{\bf 216}&337&$\pm$&200&{\bf 419}&$\pm$&
{\bf 100}&{\bf}&&{\bf }&&{\bf 2329}&\\
{\bf 48}&b&0.367&197&6&$\pm$&7&&&&{\bf 104}&$\pm$&{\bf 34}&-23&$\pm$&82
&{\bf 460}&$\pm$&{\bf 85}&{\bf 829}&$\pm$&{\bf 113}
&{\bf 739}&$\pm$&{\bf 239}&{\bf 554}&$\pm$&{\bf 142}&
{\bf 550}&$\pm$&{\bf 165}&{\bf }&&{\bf }&&{\bf 740}&\\[0.9ex]
{\bf 20}&b&0.174&31&-17&$\pm$&4&-2&$\pm$&12&-74&$\pm$&35&-73&$\pm$&28
&{\bf 97}&$\pm$&{\bf 26}&140&$\pm$&92&&&
&{\bf 167}&$\pm$&{\bf 17}&&&&0.0&$\pm$&27.6&&{\bf }&\\
{\bf 273}&c&0.158&35&&&&&&&&&&&&
&{\bf 1124}&$\pm$&{\bf 86}&{\bf 1348}&$\pm$&{\bf 68}
&{\bf 1546}&$\pm$&{\bf 94}&{\bf 1292}&$\pm$&{\bf 21}&
{\bf 1056}&$\pm$&{\bf 75}&{\bf }&&{\bf }&&{\bf 2060}&\\[0.9ex]
{\bf 295}&c&0.461&513&{\bf 16}&$\pm$&{\bf 4}&-1&$\pm$&9&42&$\pm$&64&-218&$\pm$&122
&{\bf 157}&$\pm$&{\bf 26}&{\bf 155}&$\pm$&{\bf 19}
&&&&106&$\pm$&37&&&&{\bf 19.1}&$\pm$&{\bf 0.9}&&{\bf }&\\  
{\bf 47}&b&0.425&135&0&$\pm$&3&&&&27&$\pm$&13&-22&$\pm$&15
&{\bf 103}&$\pm$&{\bf 30}&{\bf 169}&$\pm$&{\bf 53}
&&&&{\bf 164}&$\pm$&{\bf 25}&&&&{\bf }&&{\bf }&&{\bf 180}&\\[0.9ex] 
427.1&c&0.572&274&1&$\pm$&16&0&$\pm$&5&-25&$\pm$&38&36&$\pm$&25
&37&$\pm$&45&72&$\pm$&68&&&&-1867&$\pm$&60&&&&0.7&$\pm$&0.7&&{\bf }&\\
{\bf 380}&m&0.692&984&-1&$\pm$&3&2&$\pm$&1&-8&$\pm$&7&-39&$\pm$&36
&{\bf 54}&$\pm$&{\bf 16}&{\bf 56}&$\pm$&{\bf 16}
&&&&{\bf 83}&$\pm$&{\bf 20}&&&&{\bf 514.9}&$\pm$&{\bf 2.4}&&{\bf }&\\[0.9ex]
{\bf 325.0}&c&0.860&452&17&$\pm$&13&3&$\pm$&6&-29&$\pm$&25&-84&$\pm$&27
&{\bf 160}&$\pm$&{\bf 52}&{\bf 140}&$\pm$&{\bf 19}
&&&&27&$\pm$&42&&&&0.4&$\pm$&0.6&&{\bf }&\\
138&m&0.759&467&&&&&&&8&$\pm$&9&56&$\pm$&69
&13&$\pm$&42&25&$\pm$&37&
&&&27&$\pm$&39&&&&{\bf }&&{\bf }&&{\bf }&\\[0.9ex] 
280&c&0.996&1007&-3&$\pm$&1&-1&$\pm$&4&26&$\pm$&22&-10&$\pm$&68
&66&$\pm$&31&58&$\pm$&63&
&&&34&$\pm$&77&&&&{\bf 14.7}&$\pm$&{\bf 1.0}&&{\bf }&\\
{\bf 286}&c&0.849&699&6&$\pm$&3&&&&41&$\pm$&31&-1&$\pm$&39
&{\bf 142}&$\pm$&{\bf 34}&{\bf 126}&$\pm$&{\bf 39}
&&&&35&$\pm$&62&&&&{\bf 233.4}&$\pm$&{\bf 3.5}&&{\bf }&\\[0.9ex]
{\bf 368}&c&1.131&826&{\bf 6}&$\pm$&{\bf 2}&{\bf 8}&$\pm$&{\bf 2}&22&$\pm$&40&-52&$\pm$&54
&{\bf 94}&$\pm$&{\bf 24}&72&$\pm$&48
&&&&80&$\pm$&56&&&&0.14&$\pm$&0.6&&{\bf }&\\
{\bf 309.1}&c&0.905&479&{\bf 22}&$\pm$&{\bf 5}&13&$\pm$&13&18&$\pm$&62&-85&$\pm$&162
&{\bf 100}&$\pm$&{\bf 10}&-14&$\pm$&48
&&&&272&$\pm$&91&&&&{\bf 266.6}&$\pm$&{\bf 1.7}&&{\bf }&\\[0.9ex]
324&c&1.206&1127&3&$\pm$&2&0&$\pm$&1&0&$\pm$&11&37&$\pm$&194
&-29&$\pm$&34&53&$\pm$&77
&&&&236&$\pm$&79&&&&0.5&$\pm$&0.6&&{\bf }&\\
287&c&1.055&811&-5&$\pm$&6&&&&20&$\pm$&23&34&$\pm$&42
&-33&$\pm$&52&-89&$\pm$&26
&&&&&&&&&&{\bf 81.2}&$\pm$&{\bf 1.4}&&{\bf }&\\[0.9ex] 
{\bf 65}&c&1.176&1015&{\bf 10}&$\pm$&{\bf 2}&&&&7&$\pm$&41&-10&$\pm$&20
&{\bf 129}&$\pm$&{\bf 18}&-19&$\pm$&43&
&&&38&$\pm$&33&&&&0.0&$\pm$&8.9&&{\bf }&\\
{\bf 298}&c&1.436&5528&1&$\pm$&4&&&&{\bf 23}&$\pm$&{\bf 4}&{\bf 29}&$\pm$&{\bf 9}
&{\bf 184}&$\pm$&{\bf 24}&{\bf 213}&$\pm$&{\bf 39}
&&&&{\bf 243}&$\pm$&{\bf 59}&&&&{\bf 14.1}&$\pm$&{\bf 0.8}&&{\bf 218}&\\[0.9ex]
437&c&1.480&1853&1&$\pm$&7&-2&$\pm$&4&45&$\pm$&34&-23&$\pm$&46
&-31&$\pm$&42&5&$\pm$&82
&&&&-6&$\pm$&63&&&&{\bf }&&{\bf }&&{\bf }&\\
9&c&2.012&5543&7&$\pm$&3&&&&27&$\pm$&14&9&$\pm$&75
&-16&$\pm$&48&-68&$\pm$&82
&&&&-148&$\pm$&109&&&&{\bf }&&{\bf }&&{\bf }&\\[0.9ex]

\hline
\end{tabular}
\end{table*}

\section{Results}

\subsection{Spectral energy distributions}

The ISOPHOT flux measurements of the 10 pairs in our sample are
listed in Table 1. Sources which have been detected at least at the
$3\sigma$ level have their 3C number highlighted in bold face. 
In addition, new IRAM observations at 1300\,$\mu$m have been obtained
for 11 sources, seven of which have been detected. Four of the quasars
(3C\,47, 3C\,48, 3C\,273, 3C\,298) and the nearby radio galaxy Cygnus
A (3C\,405) had already been detected by IRAS (see 60\,$\mu$m fluxes 
in the last column). Leaving aside the variable quasar 3C\,273 (see
below) the ISO fluxes are on average $(25\pm20)\%$ fainter than the
IRAS fluxes.  With eight new detections in the FIR, we
increased the detection rate by more than twofold compared to IRAS.
However, seven of 20 sources (3 quasars and 4 radio galaxies) remain
undetected at all wavelengths to the limits of the ISO observations.

The measured spectral energy distributions (SEDs) are shown in Fig.\,5
supplemented by data from the literature.  They reveal signatures of
two emission components: \\ 
(A) {\it Synchrotron emission:} For one of
the quasars (3C\,380) the ISO photometry and the mm flux can be fitted
by a power-law $\nu^{-0.75}$ which joins smoothly into the radio
spectrum (near minimum) of this highly variable source.  Obviously,
the mid- to far-infrared SED is totally dominated by synchrotron
emission from the radio core which exhibits the flattest radio
spectrum in our sample. Note that the level of the synchrotron
power-law is so high that even a luminous (10$^{\rm 11}$L$_{\odot}$)
dust component could easily be hidden underneath.

In 1983, at the epoch of the IRAS observations, the mid- to
far-infrared SED of 3C\,273 was entirely dominated by synchrotron
emission. However, at the time of the ISO measurments in July 1996 the
60 to 100\,$\mu$m fluxes of 3C\,273 were about 50\% lower than
measured by IRAS. This low synchrotron activity is consistent with the
results of monitoring the source at other wavelengths  (see Fig.1 in
Stevens {\it et al.} 1998 and 1.3 and 3.3\,mm monitoring with IRAM by Hans
Ungerechts, private communication). The variability alone indicates that the
FIR flux of 3C\,273 contains a considerable synchrotron contribution,
at least during the phases of high flux, like in 1983.  
Using the mm fluxes in 1996 and a synchrotron spectrum $\sim
\nu^{-1.1}$ as  reference (see Fig. 5), we find that only at the
longest wavelengths (170, 180\,$\mu$m) the ISO measurements are
consistent with a pure synchrotron power-law, while the flux values
between 60 and 120\,$\mu$m straddle significantly above that
power-law. We attribute this to the presence of thermal dust emission,
which only can  be detected in 3C\,273 when the synchrotron core is in
a low state of activity. \\
(B) {\it Thermal dust emission:} In all detected radio galaxies and at
least four quasars the mid- to far-infrared SED exhibits a ``bump''
well in excess of the synchrotron component. We interpret this as
signature of emission from dust with temperatures between 30 and
several hundred Kelvin as is commonly seen in the infrared SEDs of
quasars ({\it e.g.} Haas {\it et al.} 2000). 
Modified blackbodies with an emissivity proportional to $\lambda^{\rm
-2}$ are fitted to the data.  In the cases with good signal/noise
ratio the dust emission has to be fitted by more than one single
blackbody of a certain temperature (see Fig.\,5).

Fortunately, the low synchrotron activity of 3C\,273 at the time of
the ISO observations enabled the detection of a thermal contribution to the
FIR spectrum.  In fact, the ISO FIR spectrum between 60 and
100\,$\mu$m with values around 1100 and 1300 mJy lies well above the
extrapolation of the mm-submm spectrum towards the IR: F$_{\rm \nu}\,=
354\,{\rm mJy} \times\,(\nu/\nu_{60})^{-1.1}$ (dashed line in
Fig.\,5).  We take this excess as an indication of the presence of a 
thermal bump in 3C\,273 which contributes roughly 500 mJy at both 60
and 100\,$\mu$m.

Note that in all seven cases for which we derive only upper limits, these
limits still allow for the presence of a stong thermal bump well 
above the synchrotron component.


\begin{figure*}
\begin{center}
\resizebox{7.5cm}{!}{\includegraphics {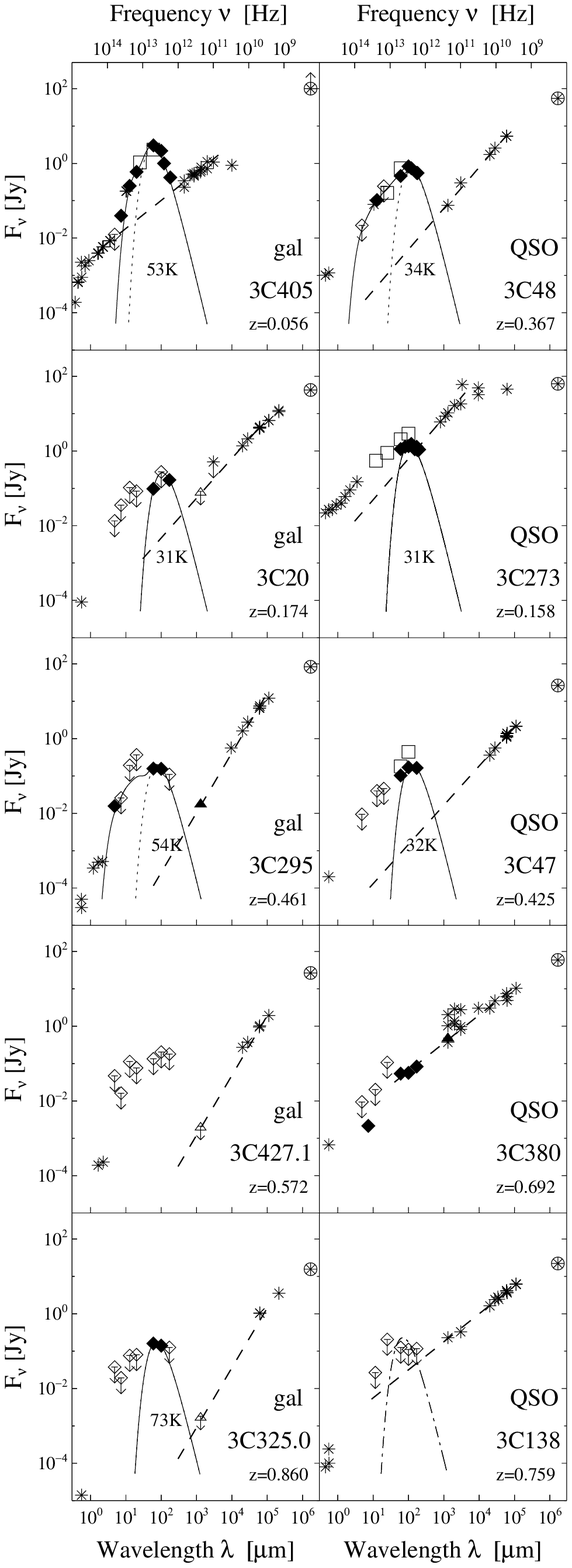} }
\hspace*{1cm}
\resizebox{7.5cm}{!}{\includegraphics {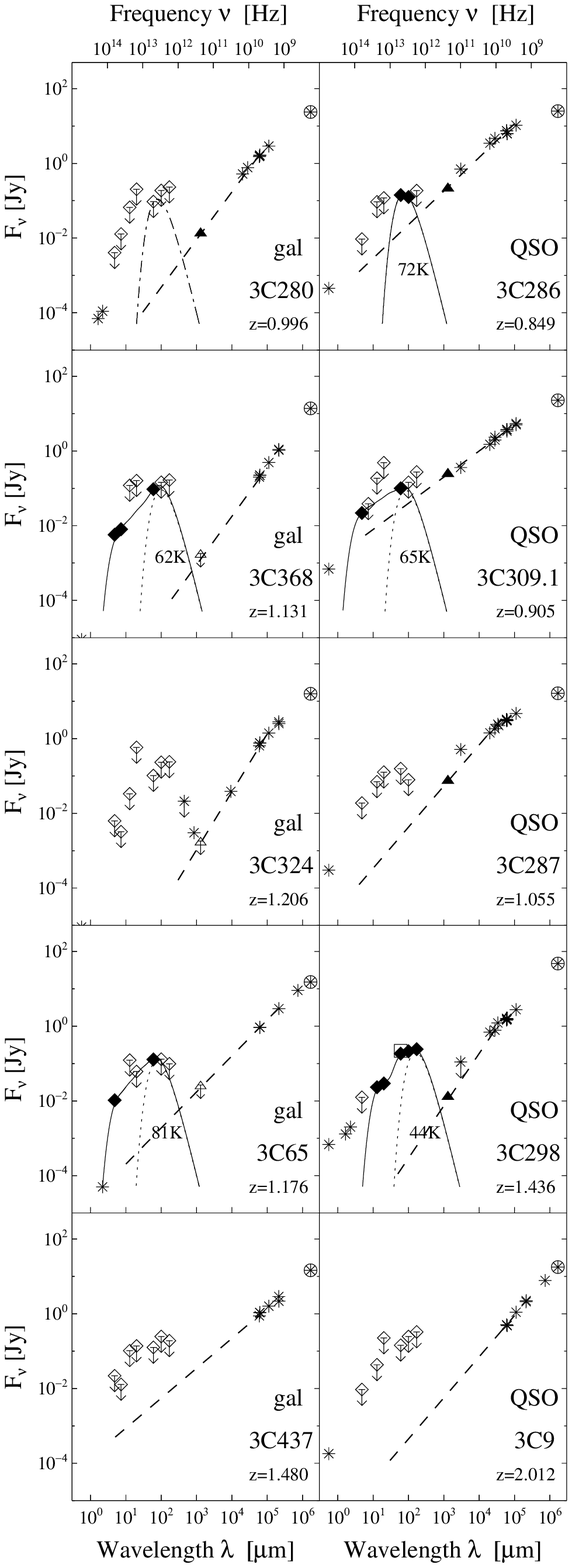} }
\end{center}
\caption[]{ Observed spectral energy distributions of the pairs of 3C
radio galaxies and quasars ordered according to the redshift of the
galaxy.  Filled diamonds refer to our ISOPHOT detections, filled
triangles to IRAM measurements, while open diamonds indicate $3\sigma$
upper limits. IRAS detections are shown by open squares. Stars give
flux measurements at optical and radio wavelengths which refer to the
central source. The exeptions are the {\it circled stars} which
represent the total flux at 178\,MHz from the 3CR catalogue.\\ The
long--dashed lines indicate the synchrotron contribution as
extrapolated from radio and mm data.  Several $\lambda^{\rm -2}$
modified blackbodies are fitted to the data (solid lines) and only the
coldest components are plotted individually (dotted lines).  Their
temperature values refer to the restframe of the object.  In cases of
non--detection of one member of a pair, the dashed--dotted line
illustrates how the blackbody of the partner would appear at
the corresponding redshift. This should be compared with the upper limits
of our observations.

}
\end{figure*}

\subsection{Pair-wise comparison of the spectral energy distributions}

As outlined above, we intend to test the unified scheme by the
comparison of radio galaxy / quasar pairs which are matched both in
extended radio luminosity and redshift. Accordingly, our results are
presented in a pair-wise manner in Fig.\,5. The pairs are sorted
according to the redshift of the radio {\it galaxy}.

Altogether, seven quasars and six radio galaxies have been detected in
at least two ISOPHOT bands. For those five pairs in which both the
radio galaxy and the quasar has been detected we indicate the
individual multi-temperature black-bodies which best fit the data. The
sole exception is 3C\,380 for which we cannot find any significant
mid- to far-infrared flux above the interpolated synchrotron
component.
Since also its partner galaxy could not be detected by ISOPHOT this pair
remains inconclusive for the purpose of testing the radio
galaxy/quasar unification hypothesis. Two more inconclusive pairs are
3C\,324 -- 3C\,287 and 3C\,437 -- 3C\,9, the latter being the pair with the
highest redshifts in our sample. In all of these four sources the upper
limits are so high that luminous thermal components are still clearly
compatible with the data. Thus, our present sample tests the
unification hypothesis out to redshifts  $z \simeq 1.2$. 

The two detected pairs at the highest redshift are 3C\,65 -- 3C\,298
and 3C\,368 -- 3C\,309.1. As will be seen below, it is of high
significance for our conclusion that in both pairs the thermal bump in
the radio galaxy and in the quasar are of comparable
strength. Moreover, all four sources show detections at mid-infrared
wavelengths revealing the existence of a hot dust component 
commonly found in AGNs.

Our sample contains two pairs (3C\,325 -- 3C\,138 and 3C\,280 -- 3C\,286) in
which only one of the partners could be detected at 60 and
100\,$\mu$m. In Fig. 5 we plot the thermal bump of the detected
partner underneath the upper limits of the undetected source. This
emphazises that the upper limits allow  for thermal emission of the
same level in {\it both} members of the pair. It is, 
therefore, not astonishing that we find a quasar with its paired galaxy
undetected as well as the opposite case.

It should be noted that in our current sample the two pairs at the
lowest redshift are the most problematic ones:\\
The pair 3C\,405 -- 3C\,48 is rather poorly matched in redshift and
extended radio power because it was formed essentially out of two loose
ends (see sample selection, Section 2, above). Nevertheless, since the radio
power of the galaxy 3C\,405 (Cygnus A, $F_{178{\rm MHz}} = 8700$\,Jy) is higher than that of the
quasar despite being at more than $6 \times$ lower redshift, the
asymmetries in this pair are balanced and should not bias the
full sample.

The pair 3C\,20 -- 3C\,273 is well matched in redshift and extended
radio power. While 3C\,20 is an archetypical FR II radio galaxy with
hot spots of very high surface brightness\footnote{The synchrotron
spectrum of one of the hot spots in 3C\,20 extends even to optical
frequencies (Meisenheimer {\it et al.} 1997).}, the quasar 3C\,273 has
rather unique properties: Its radio jet of exceptional surface
brightness shows a synchrotron spectrum reaching to optical and even
X-ray frequencies (R\"oser {\it et al.} 2000). However, even on the most
sensitive VLA maps the source appears one-sided with no trace
whatsoever of extended radio emission on the side opposite to the
jet. Most of the extended radio emission even at low frequencies
originates from the very bright hot spot at the termination point of
the radio jet. Thus, the flux at 178\,MHz does not refer to ``lobe''
emission alone. On top of that, the core displays a very strong and
highly variable synchrotron spectrum which extends to $\lambda \la
1\,\mu$m. This makes the determination of the thermal FIR emission
rather uncertain (see above). We conclude, therefore, that the unique
source 3C\,273 does not seem ideally suited to investigate the general
properties of 3CR sources.  
In general, it has to be remarked that our
sample of 3C quasars contains several flat spectrum quasars, the
beamed synchrotron spectra of which extend well into the mid- to
far-infrared wavelength region (3C\,273, 3C\,380, 3C\,309.1, and
propably 3C\,138). As long as the thermal bump from dust emission
peaks out well above the synchrotron component this does not affect
our comparison.  However, it should be noted that -- on average -- the
synchrotron component of the quasars in our sample is more than
$10\times$ brighter than that in radio galaxies.

\section{Discussion}

The present sample of 3CR radio galaxies and quasars has been selected
as pairs of objects which match as close as possible in their extended
lobe emission at very low frequency and in redshift. As outlined in
the introduction, any unified scheme which regards them as members of
the same parent population of objects seen under various aspect angles
w.r.t. the radio axis, predicts that the dust emission properties
should be identical. This obviously holds only for the sample average.
Individual pairs might differ by factors of 10 due to different dust
masses and the source geometry (see e.g. the wide variety
of the IR spectra of PG quasars, Haas {\it et al.} 2000).

There are two ways to test the unification hypothesis:
\begin{itemize}
\item[(1)] The most straightforward test simply compares the detection
statistics. According to this test the 3C radio galaxies and quasars
in our sample are indistinguishable: Among the 10 pairs we found 5 for
which both the quasar and the radio galaxy is detected by ISOPHOT, one
with only the quasar, and one with only the radio galaxy
being detected.  In the remaining 3 pairs we could neither detect the quasar
nor the galaxy. Note that all detected sources show a significant
signal at least in one of the 60 and 100\,$\mu$m bands.

If we extend our sample by including the 4 radio galaxy -- quasar
pairs which have been observed with ISOPHOT by van Bemmel, Barthel \&
deGrauuw (2000), and have been selected and evaluated along identical
lines as ours, we end up with 7 pairs in which both the radio galaxy
and the quasar is detected, 3 pairs with only the quasar, and 1 pair
with only the radio galaxy being detected. Thus, also the extended
sample is in good statistical agreement with the expectation of the
unified scheme in its most simple form.

\end{itemize}
On the other hand, this result is in stark conflict with the results
derived from IRAS (Heckman {\it et al.} 1992, Hes {\it et al.} 1995) which
indicate that radio galaxies are about $4\times$ fainter at 60\,$\mu$m
than quasars. We therefore perform a second test:

\begin{itemize}
\item[(2)] If the unified scheme is correct, not only the
detection statistics (which might be misleading due to varying detection
limits, see above) but also the emitted FIR power should be the same
for objects of identical core luminosity. If we measure the core power
by the kinetic energy which is channeled via the jets into the lobes
(that is by the radio power $\nu F_\nu$ at $\nu = 178$MHz) the ratio
of the dust to radio emission: $R_{dr} \equiv \nu F_\nu {\rm (FIR)} /
\nu F_\nu {\rm (178MHz)} $ should be identical for galaxies and
quasars.  

The top panel of Fig.\,6 displays this ratio as a function of redshift
for our sample.  All detected quasars cluster around values of $R_{dr}
\simeq 150$ with a total spread of less than a factor of three
(exception: 3C\,380 in which we only see a synchrotron component).
If we assume that only a tiny fraction of the kinetic
power in the jets (on the order of a few percent, see Rawlings \&
Saunders 1991 and Meisenheimer {\it et al.} 1997) is converted into
synchrotron radiation at 178 MHz, we find that the kinetic jet power
and the luminosity emitted by dust closely agree.

In general, we cannot find any remarkable differences between the
distribution of quasars and galaxies -- again in agreement with the
unified scheme. The only exception is Cygnus A at very low redshift
and with unusually high radio power (Barthel \& Arnaud 1996) showing
$R_{dr} < 10$.  Although the total spread of about a factor of ten in
the $R_{dr}$ of radio galaxies would be consistent with a spread in FIR
luminosities similar to that found in PG quasars (Haas {\it et al.} 2000), we note a trend with
redshift in the sense that at redshift $z < 0.5$ the typical $R_{dr}$
seems to be 4 times lower than at $z \simeq 1$. We identify two
groups: Radio galaxies at $z < 0.5$ with $R_{dr} \la 50$ (sole
exception: 3C\,67 from the van Bemmel {\it et al.}  sample) and those at $z
\ga 0.8$ with $R_{dr} \ga 150$ which is also typical for the quasars. 
The alternative hypothesis that
$R_{dr}$ is primarily correlated with radio power is not supported
(see bottom panel of Fig.\,6) since {\it e.g.} the radio galaxies
3C\,405, 3C\,295, 3C\,368, 3C\,325 and 3C\,65 cover only a factor of 2
in radio power while their $R_{dr}$ varies by more than a factor of 30.
\end{itemize}
The fact that $R_{dr}$ correlates better with redshift than with
luminosity
can be understood if the thermal power of a radio source is primarily
controlled by the fueling rate which should be higher at large
redshifts when violent mergers have been more frequent than today. The
dependance of $R_{dr}$ on redshift also provides a natural explanation
for the apparent discrepancy between our result (1) and the IRAS
results: Counting on the superior sensitivity of the pointed ISOPHOT
observations we deliberately biased our sample towards the
high-redshift galaxies and quasars in the 3CR sample, while most IRAS
samples are dominated by objects at moderate ($z < 0.8$) or low
redshift ($ z < 0.3$, Heckman {\it et al.} 1994).

\begin{figure}
\hspace{-0.65cm}
\centering
\resizebox{8.cm}{!}{\includegraphics {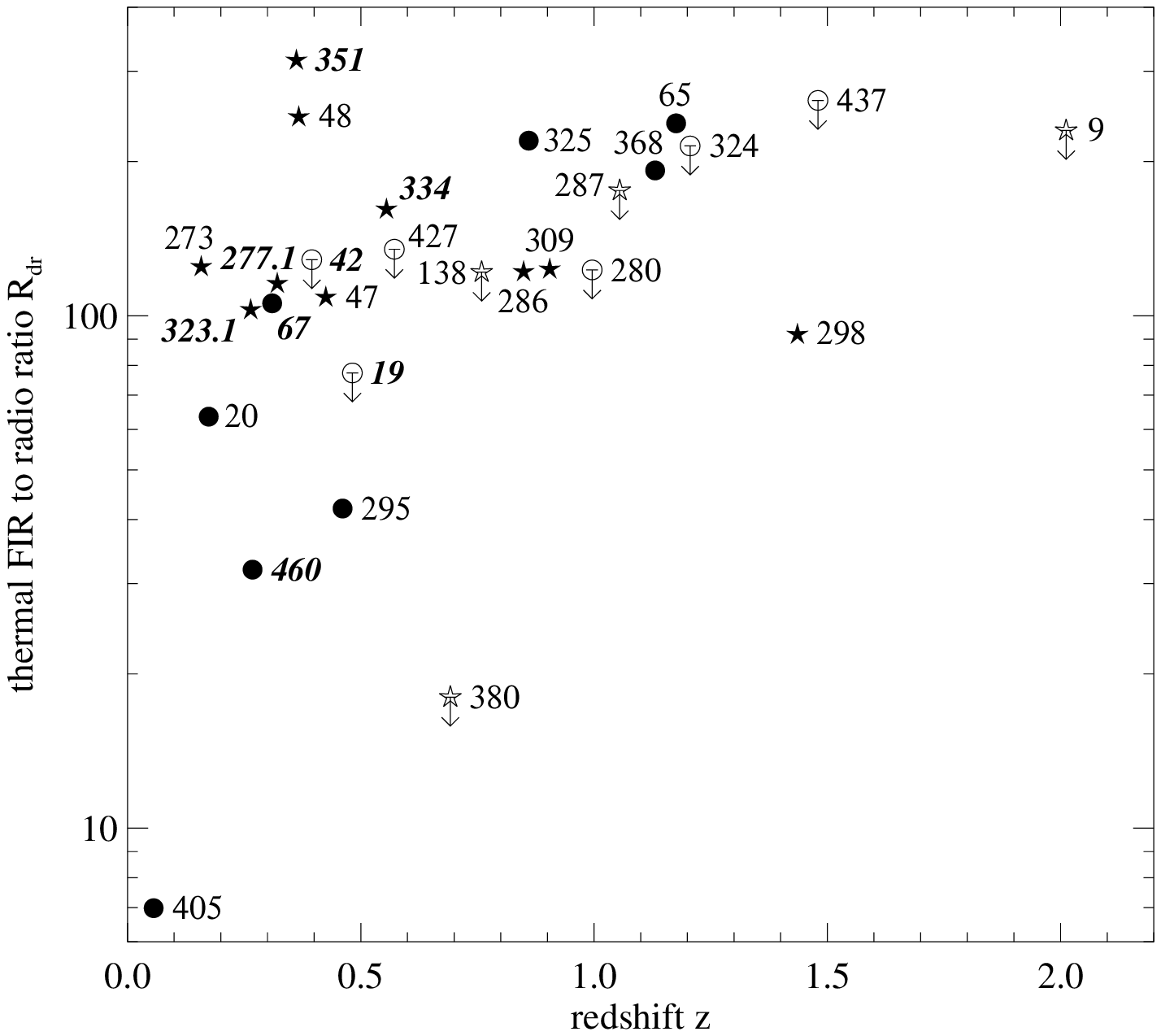} }
\resizebox{8.cm}{!}{\includegraphics {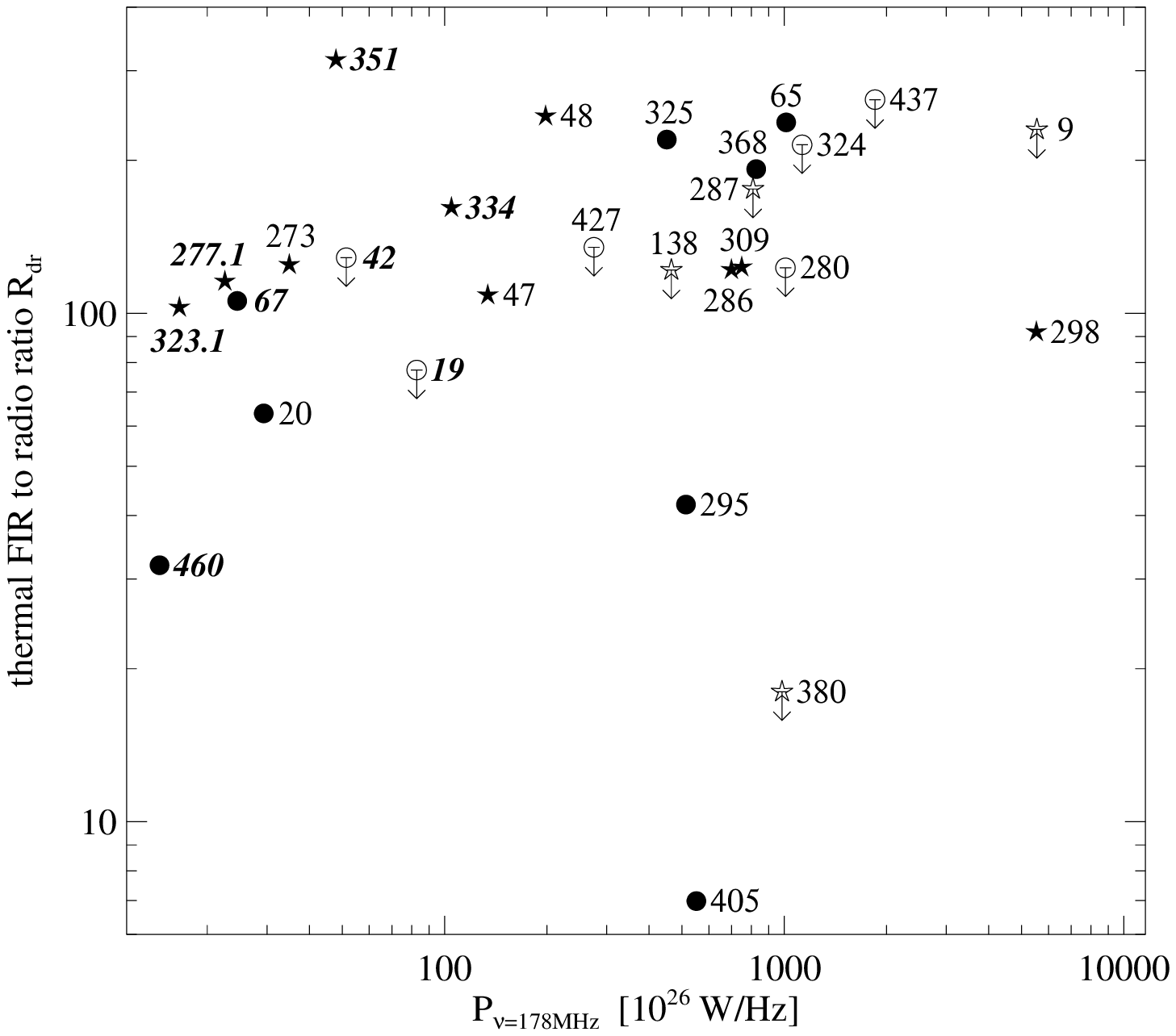} }
\caption[]{ Ratio $R_{dr}$ of thermal dust power $\nu F_\nu$
(averaged over 60 and 100\,$\mu$m) to radio power $\nu F_\nu$ at 178
MHz. Sources from van Bemmel {\it et al.} (2000)  have their 3C number shown in
italics. \\
{\it Top panel:} $R_{dr}$ as function of $z$. {\it Bottom
panel:} $R_{dr}$ as function of spectral radio power $P_\nu$ at 178\,MHz.  }
\end{figure}

The redshift dependance of the $R_{dr}$ in 3C radio galaxies is in
intriguing coincidence with the statistical properties of the
so-called ``alignment effect'' between the orientation of extended
emission line regions (EELRs) around luminous radio galaxies and
their radio axis (Chambers {\it et al.} 1987, McCarthy {\it et al.} 1987): Good to
perfect alignments ($\Delta PA \equiv |PA{\rm (EELR)} - PA{\rm
(radio)}| < 20\degr$) are commonly found in 3C galaxies above $z=0.6$
while low redshift radio galaxies show hardly any significant
alignment (Baum {\it et al.} 1988).
The most popular explanation for this finding assumes that the aligned
radio galaxies at high redshift contain a quasar core which emits its
ionizing UV predominantly along the radio axis. In this interpretation
the alignment effect could be used as further independent evidence for
the unified scheme. In fact, three of the radio galaxies in our sample
(3C\,280, 3C\,324, 3C\,368) have long been known for their perfectly
aligned and very extended emission line region ($D > 20$\,kpc).  In
one of these sources, 3C\,368, we find a strong mid- and far-infrared
source with a luminosity of $2 \times 10^{13} {\rm L}_{\sun}$; in
3C\,324 we can only determine a rather high upper limit which is still
compatible with a luminous quasar.
Only 3C\, 280 (which has the smallest EELR of the three)
seems to lie at the low end of the $R_{dr}$ distribution. 

Additionally, our sample contains one radio galaxy, 3C\,65, which has
been suspected to contain a hidden quasar (Lacy {\it et al.} 1995, Simpson
{\it et al.} 1999). Indeed, together with 3C\,351 and the archetypical
quasar 3C\,48, this radio galaxy exhibits the largest $R_{dr}$ in our
sample.  From the three radio galaxies at $z > 0.8$ detected with
ISOPHOT, only 3C\,325 has not been suspected of containing a quasar
core on the basis of optical or NIR observations. Unfortunately, our
MIR sensitivity for this source does not suffice to draw any
conclusions about the presence of warm dust emitting at $\lambda <
30\,\mu$m. The same is true for the low-redshift galaxies 3C\,20 and
3C\,460. On the basis of the present data one cannot exclude the
possibility that part of the FIR luminosity is generated by a
starburst in the radio galaxy without direct connection to the active
core.  In fact, one might speculate what fraction of the FIR radiation
(from dust at temperature $< 50$\,K) in the other sources of our
sample could be attributed to starburst activity in the host
galaxy. Better sampled mid-infrared spectral energy distributions or
spectroscopic data would be required to settle this issue. If
off-nuclear sturbursts play indeed an important role in the overall
FIR emission at $\lambda \ge 60\,\mu$m, the significance of the ``FIR
test'' for unified schemes could be considerably weakened.


\section{Conclusions}

New ISOPHOT observations of 10 pairs of radio galaxies and quasars
from the 3CR Catalogue between 5 and 180\,$\mu$m cannot find
any statistical difference between the detection rate of radio galaxies and quasars.
This is in perfect agreement with the unified scheme proposed
by Barthel (1989) which postulates that radio galaxies and quasars
are intrinsically identical objects only differing by the angle
between the radio axis and the line of sight.

In an attempt to quantify the unification further, we
introduce the ratio
of the dust to radio emission: $R_{dr} \equiv \nu F_\nu {\rm (FIR)} /
\nu F_\nu {\rm (178MHz)} $.  While for quasars $R_{dr}$ scatters by a
factor of a few for all redshifts and radio powers, we find a clear
trend with redshift for the radio galaxies:

At low redshifts $z < 0.8$ hardly any of the galaxies reach the high
dust-to-radio luminosity ratios $R_{dr} \simeq 150 $ typical for the
quasars. This is roughly consistent with the conclusion drawn from the
IRAS results that radio galaxies are about 4$\times$ weaker FIR
sources than quasars. At the high redshift end $z > 0.8$, however,
several radio galaxies show thermal dust emission in the mid- and
far-infrared which is in every aspect comparable to that of the IR
brightest quasars. Beyond any doubt, these sources contain a hidden
quasar and thus favour the unified scheme. Since our ISOPHOT sample
contains a much higher fraction of high redshift radio galaxies than
any other radio/galaxy sample observed before in the FIR, we find for
the first time a good agreement with the prediction of the unified
scheme.

Interestingly, there exists independent evidence from the ``alignment
effect'' between the extended emission line region around radio
galaxies and the radio axis that many of the most luminous radio
galaxies at $z \ge 0.6$ contain a quasar.

On the basis of the current knowledge, we therefore propose
the following refinement of Barthel's unified scheme for luminous radio-sources:
At $z \ga 0.7$ every galaxy with extended radio luminosity $\nu F_\nu
> 3 \times 10^{36}$ W contains a quasar. At low redshift, even among
the brightest radio galaxies there exist some sources with -- compared
to the radio power -- under-luminous thermal cores. Maybe these
``true'' radio galaxies suffer from the exhaustion of the
circum-nuclear fuel which governs the accretion rate onto the central
black hole.  The idea that many of today's massive black holes are
starving also finds support in the rapid decline of the density of
optically selected quasars between $z = 3$ and the local universe.

\acknowledgements It is a pleasure for us to thank Hans Ungerechts at
IRAM for providing us with the monitoring fluxes of 3C273.  For
literature search and photometry NED and SIMBAD data bases were used.
The development and operation of ISOPHOT and the Postoperation Phase
are supported by MPIA and funds from Deutsches Zentrum f\"ur Luft-und
Raumfahrt (DLR, formerly DARA). The ISOPHOT Data Centre is supported
by DLR with funds from Bundesministerium f\"ur Bildung und Forschung,
grant No. 50\,QI98013.  The authors are responsible for the contents of
this paper. The Interactive Analysis tool PIA is a joint
development by the ESA Astrophysics Division and the ISOPHOT
consortium. We thank an anonymous referee for valuable comments which
greatly improved the original manuscript.

\end{document}